\documentclass{aa}  
\bibpunct{(}{)}{;}{a}{}{,} 
\begin{document}

   \title{Parallax measurements of  cool brown dwarfs\thanks{Based on observations taken with Omega-2000 at the 3.5 m telescope at the Centro Astron\'omico Hispano Alem\'an (CAHA) at Calar Alto, operated by the Max Planck Institut f\"ur Astronomie and the Instituto de Astrof{\'{i}}sica de Andaluc{\'{i}}a (CSIC).}}

   \subtitle{}

   \author{E. Manjavacas 
          \inst{1}, 
          B. Goldman\inst{1},
	  S. Reffert\inst{2}
	  \and T. Henning\inst{1}
          }

	\institute{Max Planck Institute f\"ur Astronomie.
              K\"onigstuhl, 17. D-69117 Heidelberg, Germany\\
              \email{[manjavacas;goldman;henning]@mpia.de}
         \and
             Landessternwarte
	      K\"onigstuhl 12, D-69117 Heidelberg, Germany\\
             \email{sreffert@lsw.uni-heidelberg.de}
             }

   \date{Received 18 April 2013; accepted October 2013}

 


 

  \abstract
   {Accurate parallax measurements allow us to determine physical properties of brown dwarfs and help us  constrain evolutionary and atmospheric models,  break  age-mass degeneracy, and reveal unresolved binaries.}
   {We measured absolute trigonometric parallaxes and proper motions  of six cool brown dwarfs using background galaxies to establish an absolute reference frame. We derive the absolute $J$-band magnitude. The six T brown dwarfs in our sample have spectral types between T2.5 and T8 and magnitudes between 13.9 and 18.0 in 2MASS (Two Micron All Sky Survey) with photometric distances below 25~pc. }
   {The observations were taken in the $J$-band with the Omega-2000 camera on the 3.5~m telescope at Calar Alto  during a time period of  27 months  between March 2011 and June 2013. The number of epochs varied between 11 and 12 depending on the object. The reduction of the astrometric measurements was carried out with respect to the field stars. The relative parallax and proper motions were transformed into absolute measurements using the background galaxies in our fields. }
   {We obtained absolute parallaxes for our six brown dwarfs with a precision between 3 and 6~mas. We compared our results in a color-magnitude diagram with other brown dwarfs with determined parallax and with the BT-Settl 2012 atmospheric models. For four of the six  targets, we found a good agreement in luminosity with objects of similar spectral types. We obtained an improved accuracy in the parallaxes and proper motions in comparison to previous works. The object 2MASS~J11061197+2754225 is more than 1~mag overluminous in all bands, which point to binarity or high order multiplicity. }
 {}

   \maketitle
%

\section{Introduction}\label{introduction}

Since the discovery of the first brown dwarfs (BDs) (GD165 by \citealt{Becklin_Zuckerman}, PPl~15 by \citealt{Basri},  Teide1 by \citealt{Rebolo} and  Gliese229B by \citealt{Nakajima}),  more than 1000 L and T type brown dwarfs have been discovered\footnote{\textit www.dwarfarchives.org}. 

During their evolution BDs  cool and dim. During their life, they change their spectral type  because the effective temperature ($T_{\rm eff}$) decreases with the exception of BDs in the L/T transition in which the $T_{\rm eff}$ is roughly constant. For a given  BD with a given temperature, the interval of masses and ages that the BD could have is very wide, so that the age and the mass are degenerate. The  $T_{\rm eff}$ is the physical parameter that drives the major changes in the observable photometric and spectroscopic features of  brown dwarfs. However, among the known brown dwarfs emerge outliers, which show that secondary parameters are also responsible for  brown dwarf properties, such as  gravity,  metallicity, and cloud variability \citep{Burrows, Burgasser2006b, Liu2007}.

Accurate measurements of distances allow us to determine physical properties of these objects, including luminosities or absolute fluxes to check atmospheric models, temperatures, space motions, and space densities.

Since our goal is to derive parallaxes with a high accuracy (5-10\%), we constrained our sample to objects with photometric distances up to 25~pc. Our sample of objects is located at spectrophotometric distances of 10$-$25~pc with magnitudes in $J$ between 13.9 and 18.0, which are suitable for astrometry.

Since the first parallax programs for brown dwarfs began with  \citet{Dahn}, \citet{Vrba}, and \citet{Tinney},  the relationship between the color and magnitude of BDs has been studied \citep{Burgasser2008, Schilbach, Marocco,  Faherty, Dupuy} among others. One of the most significant results of these studies is the large dispersion in luminosity for objects with similar spectral types \citep{Faherty}, which shows the importance of other factors, such as gravity, metallicity, sedimentation, and binarity (\citealt{Tsuji};  \citealt{Burrows};  \citealt{Saumon-Marley}). Increasing the number of cool brown dwarfs with accurate distance measurements allows us to understand the variation in the color magnitude and H-R diagrams, as we can determine the luminosity more accurately. Also, the $J$-band bump in the color-magnitude diagram, a brightening observed in the J band for brown dwarfs with spectral types between T1 and T5, is not well understood (\citealt{Burgasser2002}, \citealt{Tinney} and  \citealt{Vrba}). There are still few objects with parallaxes in the L/T transition, which prevent the progress of understanding brown dwarf evolution.

In this paper, we report the absolute parallaxes and the absolute proper motions of six T ultra cool dwarfs with spectral types between T2.5 and T8 and spectrophotometric distances between 10$-$25~pc.

In Section \ref{data}, we explain how our selection criteria are used to select our targets and how the observations were carried out. In Section \ref{Analysis}, we describe the data reduction: how the astrometry was performed, the estimation of the differential chromatic refraction (DCR), and the calculation of the relative and absolute parallax and proper motions. 
In Section \ref{results}, we compare our results to those of other previous studies, which included our object. Finally, we present our conclusions in Section \ref{conclusions}.


\section{Observations}\label{data}

Initially, we selected eigth T dwarfs with expected spectrophotometric distances smaller than 25~pc. Spectrophotometric distances of our targets were estimated using the relation given by \citet{Goldman}. Only objects brighter than 18~mag in the $J$ band were selected, so that good signal-to-noise (S/N) observations could be obtained in a reasonable amount of time. Two targets were discarded later from the list, since they were located in  fields with bright stars nearby, which compromised the accuracy of the astrometry. Finally, all the targets had to be observable  most part of the year to have a better coverage of the parallax ellipse. In Table~1, we present an overview of our targets, which provide spectral types and derived JHK  photometry in the MKO (Mauna Kea Observatories) system \citep{Stephens} for targets with 2MASS photometry. For the rest, we provide UKIDSS (UKIRT Infrared Deep Sky Survey) photometry, which is similar to MKO  \citep{Hewett}. For the target ULAS~J232035.28+144829.8, we provide the photometry from \citet{Murray}, which is also similar to MKO. In the last column, we add  references to the discovery papers.

\begin{table*}
  \caption{This table gives the sample of objects with their spectral types, photometry and references for our targets.}
\label{targets}
\begin{center}
\begin{tabular}{l c c c c c c c l}
 \hline
 \hline
Name & SpT & $J$ [mag]  & $H$ [mag]  & $K_{s}$ [mag]  & ref. objects & exS & epochs & Reference  \\
 \hline
2MASS~J11061197+2754225  & T$2.5$ (1) & 14.96$\pm$0.04 & 14.20$\pm$0.05 & 13.84$\pm$0.05    &  96    & 49  & 10 & (1) \\
ULAS J130217.21+130851.2 & T$8$ (5) & 18.11$\pm$0.04 & 18.60$\pm$0.06 & 18.28$\pm$0.03      &  247   & 68 & 12 & (2) \\
ULAS~J141756.22+133045.8 & T$5.5$ (3)  & 16.77$\pm$0.01 & 17.00$\pm$0.03 & 17.00$\pm$0.04   &  77    & 99  & 10 & (3)\\
2MASS~J22541892+3123498 & T$4$ (4) & 15.32$\pm$0.05 &15.06$\pm$0.08 & 14.99$\pm$0.15        & 644    & 125  & 10 &(4) \\
ULAS~J232035.28+144829.8  & T$6$ (6) & 16.79$\pm$0.02 & 17.14$\pm$0.04 & 17.40$\pm$0.03     &  298   & 46 & 12 & (3) \\
ULAS~J232123.79+135454.9 & T$7.5$ (2) & 16.69$\pm$0.03 & 17.09$\pm$0.06 & 17.36$\pm$0.10    & 278    & 73 & 11 & (3)\\ 
\hline

\end{tabular}
\tablebib{
(1)~\citet{Looper}; (2) \citet{Burningham2}; (3) \citet{Scholz}; (4) \citet{Burgasser};
 (5) \citet{Cushing}; (6) \citet{Murray}.
}
\end{center}
\end{table*}


Images were taken with  the near-IR camera Omega-2000 on the 3.5 m telescope at  Calar Alto in $J$ band.  Omega-2000 is a prime-focus, near-IR camera with a wide field that uses a 2k x 2k  focal plane array with  sensitivity from the {\textit z} band to the {\textit K} band. The camera provides a 15.4' x 15.4' field of view with a resolution of 0.45"/pixel. The wide field of Omega-2000 allows us to convert from relative parallaxes and proper motions to absolute values using galaxies (between 50 and 130 galaxies in each field).
The astrometric observations were taken in service mode in the $J$ band. In all the cases, 15 single frames with exposure times of 60s were taken with dithering. 

Our observations have been taken between March 2011 and June 2013 in 24 epochs. The baseline varies from 23 to 27 months.  In the case of the object ULAS~J232035.28+144829.8, we also used one observation, which was taken  in July 2009 with Omega-2000 in the methane filter; therefore, the baseline for this object is almost four years.  The average seeing on Calar Alto during our observations was around 1.0". We typically observed  at 1-2~hr from the meridian, so that  DCR  might be significant (\citealt{Stone}, \citealt{Pravdo}). Further details about the estimation of the DCR corrections are given in Section \ref{DCR_correction}.

Dark frames and sky flats were taken every evening, and the bad pixel mask was derived from the dark current analysis by an appropriate cut in the goodness-of-fit of the linear relation between dark current and exposure time. The fifteen individual raw images were corrected using flats, darks and bad pixels mask frames. To perform the reduction of the raw images, we used the MPIA Omega-2000 pipeline, which runs under MIDAS. The outputs are single, calibrated images for each epoch. Before performing the analysis of our images, we stacked the fifteen single exposures  to get a better S/N.  The stacking process was carried out using the algorithm explained in \citet{Mutchler_Fruchter} and \citet{Fruchter_Hook}. The single images are slightly shifted, so they must be previously aligned. These images are aligned in a world coordinate system, so that corresponding astronomical objects are stacked on top of each other. A median image  of the shifted frames is extracted. Since the images were shifted with different offsets, only the central area overlaps for all frames. The outer regions, where individual images do not contribute are marked and left blank. To check if the stacking actually improved the final value for the parallax and proper motions, we repeated the analysis using the fifteen single frames and the result obtained was less accurate than using  stacked images. Therefore, we used the stacked frames for the final analysis.


\section{Analysis}\label{Analysis}

\subsection{Astrometry}\label{astrometry}

We obtained positional measurements for all of the sources in each field from {\textit SExtractor}  \citep{Bertin-Arnouts} using the parameter XWIN\_IMAGE.   {\textit SExtractor} determines the background and identifies whether
pixels belong to background or to objects. The program splits up the area that is not background into
separate objects and determines the properties of each object. The output from {\textit SExtractor} is a catalog for each epoch and  field that contains  all the objects in each field, the positions with error, instrumental magnitudes and errors, instrumental fluxes and errors, and star/galaxy classification among other parameters. The errors in position provided by {\textit SExtractor} are estimated using photon statistics. This  estimate is considered to be  a lower value of the real error.

The next step to create an astrometric catalog with the objects in our fields  was to associate the detections in the multiple epochs that  belonged to a common set of objects. For that, we cross-identified stars and we matched detections in a given frame to an astrometric reference catalog. As the telescope did not provide WCS information in the image headers, we used the software from {\textit www.astrometry.net} \citep{Lang} to perform a preliminary astrometry. The reference catalog used is the USNO-B1.0 catalog  \citep{Monet}. This preliminary astrometry cannot be  better than the accuracy of the catalog, which is around 200~mas.  Then, we refined the initial guess using the software {\textit SCAMP} \citep{Bertin} by choosing the first stacked image of the first epoch for each target as reference catalog. The reference objects used to perform the astrometry were distributed uniformly in the fields. These fields contained between 70 and 650 references. A different weight was given to the high S/N reference objects and the small S/N reference objects.

We constructed the catalog of associated detections by starting with the list of detections in the first stacked image on the first epoch, and then adding detections from the next epoch by finding matches between the objects in the catalogs. We performed the match of the catalogs using the IRAF routine {\textit tables.ttools.tmatch}. This routine deletes the objects, which are not detected in all the epochs, but  mainly faint targets and stars at the edges of the fields. We used a search radius of 1" around the object from the reference catalog.

\subsection{DCR correction}\label{DCR_correction}

The DCR   effect results in astrometric shifts of the centroids (in any single-band imaging survey) because of the dependence  of the refractive index of air on the wavelength.  The reference objects and the targets have a different flux distribution in the $J$ band because they have different spectral types. Thus, their positions shift relative to one another due to different amounts of atmospheric refraction. Therefore, the DCR is a potential source
of astrometric error, and it must be estimated.

In our case, the observations were typically performed between 1 and 2~hr from the meridian. To estimate the magnitude of the resulting DCR  for the field stars and the targets, we used the formula according to \citet{Monet1992}, \citet{Stone}, \citet{Pravdo}, and \citet{Kaczmarczik}. We calculated  the DCR effect for the targets relative to the typical field stars (M dwarfs). The correction due to DCR was typically 1~mas for the relative position between target and field stars. Since our typical parallax errors are of the order of 3-6~mas (see Section \ref{correction_absolute_parallax}), it was not necessary to take the DCR into account.  We also checked the influence of an epoch taken far from the meridian in the final result for the parallax. The conclusion was that the final parallax value did not change significantly. We  discuss other astrometric error sources in detail in Section \ref{absolute_parallax_accuracy}.

\subsection{Parallaxes}\label{relparallaxes}

\subsubsection{Relative parallaxes}

For the target and each object in the field, we use the positions in each epoch  with their errors as an input to a $\chi^{2}$ fit. Errors used for the fitting are  the residuals in right ascension and declination, respectively, and have been determined iteratively. These errors include all the error sources explained in Section \ref{absolute_parallax_accuracy}. We fit the positions ($\alpha_{0}$ and $\delta_{0}$), proper motions ($\mu_{\alpha}$ and $\mu_{\delta}$), and the parallax ($\pi$). Each fit to  2 x \mbox{$N_{epochs}$} measurements had 2 x \mbox{$N_{epochs}$} -- 5 degrees of freedom.  It is important to note that this parallax is a {\textit relative parallax},  which uses  field objects in the field of the target as references.\\

The apparent trajectory of each object in the field was then fitted to an astrometric model:

\begin{equation}
\Delta\alpha(t) = \mu_{\alpha}(t - t_{0}) + \pi \cdot(p_{\alpha}(t)-p_{\alpha}(t_{0}))
\label{diff_alpha}
\end{equation}
\begin{equation}
 \Delta\delta(t) = \mu_{\delta}(t - t_{0}) + \pi \cdot(p_{\delta}(t)-p_{\delta}(t_{0})),
\label{diff_delta}
\end{equation}

\noindent where $\Delta\alpha(t)$ and $\Delta\delta(t)$ are the positional offsets with respect to the first epoch of observation at $t_{0}$; $t$ is the time; $\mu_{\alpha}$ and $\mu_{\delta}$ are the proper motion in RA and DEC, $\pi$ is the parallax; and $p_{\alpha}$ and $p_{\delta}$ are the parallax factors in RA and DEC, respectively. 

The parallax factors were computed using the Earth geocenter as obtained from the JPL DE405 solar system ephemeris. 
This model is based on the methods in the $Hipparcos$ \citep{Perryman} and $Tycho$ Catalogues  \citep{Hog}.

We present the plots of the stellar paths obtained using Equations \ref{diff_alpha} and \ref{diff_delta} in the appendix.

\subsubsection{Correction from relative to absolute parallax}\label{correction_absolute_parallax}

As mentioned before, the  parallaxes from the astrometric solution are relative to the position of the background objects chosen as references. In general, these objects were field objects that may have their own parallaxes and proper motions, so that a correction is  based upon the true parallaxes of the reference objects to convert to an absolute measurement. The field objects used as references are weighted depending on their S/N, giving more weight to the objects with better S/N. Given the large field of view of Omega-2000 (15.4'~x~15.4'),  we can find a sufficient number of extragalactic sources in all our fields. To derive the absolute parallax from the relative parallax, we used the extragalactic sources (exS) that we found in the fields of our targets. 
We searched for the extragalactic sources by setting the keyword CLASS\_STAR in {\textit SExtractor}. 
This software classifies extragalactic sources and stars in a field using neural networks, as explained in \citet{Bertin-Arnouts} by applying a method called $backpropagation$. Afterward, we determined the relative parallaxes of all the objects that were classified as extragalactic sources. We made a histogram of parallaxes and proper motions of these sources.  As  we could detect plenty of outliers in the histograms, we removed the objects with parallaxes or proper motions that were further than 3-$\sigma$ away from the median in parallaxes or proper motions. We fitted a Gaussian function to the histogram of the  parallaxes after deleting these outliers.  Finally, the relative parallax for the extragalactic sources  ($\pi_{exS}$) is the mean of the fitted Gaussian, and the error of this parallax is the error in the mean for the Gaussian fitted to our data.

 
We correct the relative parallax to the  absolute parallax  as follows: $\pi_{abs} = \pi_{rel} - \pi_{exS}$, where $\pi_{abs}$ is the absolute parallax for the object, $\pi_{rel}$ is the relative parallax of the object, and $\pi_{exS}$ is derived as described above.

We execute a similar procedure to calculate the absolute proper motions given in Table \ref{parallaxes}.

\subsection{Absolute parallax accuracy}\label{absolute_parallax_accuracy}

We have several error sources, which arise at different phases in our analysis:

\begin{itemize}

	\item Centroid errors. These errors are calculated by {\textit SExtractor} using an iterative method. The positional uncertainties  due to image centroiding in the relative position of the target  is on average $\sim$1.5~mas.

	\item Atmospheric image motions. Atmospheric effects also limit the precision of the astrometry in several different ways, such as intensity scintillation, image blurring, image motion and speckle structure.  We estimated this effect using the expression given in \citet{Lindegren}: $\epsilon^{2} \simeq 0.71R^{2/3}T^{-1}$ under the condition: $14T \gg 4300R \gg d$. The parameter $T$ is the integration time in seconds (60~s~x~15~single~images), $R$ is the diameter of the field in radians ($4.5*10^{-3}$~rad) and $d$ is the diameter of the telescope in meters (3.5~m).  The variance for the atmospheric image motion is estimated to be 5~mas per epoch.

	\item DCR effect. This error was calculated as explained in Section \ref{DCR_correction}, and it is typically 1~mas per epoch.

	\item Plate solution. Distortions  are introduced by the instrument because it is not sufficiently stable, as it is dismounted between two epochs. The program {\textit SCAMP} calculates the best astrometric solution performing a $\chi^{2}$ minimization and  calculates the distortions using the relative positions of the reference objects between the reference catalog and the catalogs of all epochs. The value of the residual distortion is between 5 and 10~mas for bright stars per epoch. The distortions due to the  images stacking are negligible. We measure up to 0.04\% distortions or 0.5 pixel distortions from center to corner \citep{Bailer-Jones}.

	\item Error in the conversion from relative to absolute parallaxes. It is calculated Section \ref{correction_absolute_parallax}. The typical value of these uncertainties is $\sim$ 1.5~mas.

\end{itemize}

\section{Results}\label{results}

The astrometric and photometric results for the six objects are compiled in Table \ref{parallaxes}. Column 1 contains the object's name. Columns 2 and 3 give the derived absolute proper motions in right ascension ($\mu_{\alpha}~$) and declination ($\mu_{\delta}~$) for the targets;  column 4 gives the relative parallax ($\pi_{rel}$). In  column 5, we provide the absolute parallax ($\pi_{abs}$). Column 6 
contains the derived distance ({\textit d}) from the absolute parallax results. Column 7 provides the photometric  distance ($d_{phot}$) in pc, using the relation published in \citet{Goldman}, and  column 8 contains the values for  $\chi^{2}$ and $N_{dof}$, which is the number of degrees of freedom.

To characterize our targets, we plot them in a color-magnitude diagram (CMD)  with the 177 L, L-T, and T brown dwarfs published by \citet{Dupuy} with magnitudes in the  MKO system. Our magnitudes were originally  in the 2MASS photometric system, so that  we use the relation published in \citet{Stephens} to transform between the 2MASS photometric system and the MKO system. The rest of the targets have UKIDSS photometry, which is similar to the MKO photometry \citep{Hewett}. For the target ULAS~2320+1448 we provide the photometry from \citet{Murray}, also similar to MKO.  We plot ($J-H$, $M_J$) and ($J-K$, $M_J$) and ($W1-W2$, $M_w1$)  using WISE photometry.

We overplot our CMD with the BT-Settl models by \citet{Allard}. The color-magnitude diagrams are shown in  Figure \ref{J-H_MJ_models}  with overplotted isochrones from \citet{Allard}. We show the stellar paths for all the objects in the Appendix.

\section{Discussion}\label{discussion}

\subsection{2MASS J11061197+2754225}\label{J1106}

This object was discovered by \citet{Looper}. It was identified using a near infrared (NIR) proper-motion survey based on multi-epoch data from the Two Micron All Sky Survey (2MASS).  It was classified as a T2.5 BD using the IRTF SpeX spectrograph in low-resolution mode with a resolution of R$\sim$150. \citet{Looper} estimated  the spectrophotometric distance as $15.5\pm1.2$ pc, which was calculated using the spectral types that were derived in the same article and by using  the \citet{Liu2006} spectral type versus magnitude relation without known binaries. \citet{Burgasser2010} proposed this target to be a strong binary candidate in a $T0.0\pm0.2$ and $T4.5\pm0.2$ system, although \citet{Looper2008} performed high angular resolution imaging with NIRC2/Keck~II, finding only one source.

The location of the object in  Figures \ref{J-H_MJ_models}  and \ref{W1-W2_SpT}  seem to give a luminosity inconsistent with the classification provided before  (\citealt{Looper}; \citealt{Kirkpatrick2010}) by using indices in the NIR. \citet{Looper} mentions  that J1106+27 is distinctly (> 1-$\sigma$) bluer in the z-J band in SDSS than all the T2 brown dwarfs in their sample. In our CMD (see Fig. \ref{J-H_MJ_models} and \ref{W1-W2_SpT}), the object is also overluminous by  $\sim$1~mag.  Due to its overluminosity, this object is a good candidate for a binary system or even high order multiplicity, although this issue should be studied in more detail using high resolution images or high resolution spectroscopy. The spectroscopic distance provided by \citet{Looper} is not compatible with the distance derived from the trigonometric absolute parallax within the error bars. The calculated distance is $20.6_{-1.2}^{+1.0}$~pc.

\subsection{ULAS J130217.21+130851.2}\label{J1302}

This object was discovered by \citet{Burningham2}, and it was classified as a T8.5 BD by using NIRI on the Gemini North Telescope and IRCS on the Subaru telescope on Mauna Kea, Hawaii. The spectral classification was based on the spectral indices given by \citet{Burgasser2006}. The object ULAS~J1302+13 was later re-typed as T8 by \citet{Cushing}.

With regard to Fig.~\ref{J-H_MJ_models}  and \ref{W1-W2_SpT}, this object is not overluminous. The value of the distance has been calculated for the first time and  is  $15.4_{-1.4}^{+1.1}$~pc.

\subsection{ULAS J141756.22+133045.8}\label{J1417}

The discovery of ULAS J141756.22+133045.8 was first published by \citet{Scholz}. It was classified as T5.5$\pm$1.0 BD and  based on colors and absolute magnitudes from UKIDSS and SDSS. The absolute proper motions were also estimated, using UKIDSS and SDSS data with a baseline of five years, but only with three different epochs.

The result supplied in \citet{Scholz} for the absolute proper motions were $\mu_{\alpha}\cos{\delta} = -76 \pm 3 $~mas/yr  and $\mu_{\delta} = 77 \pm 3 $~mas/yr. \citet{Scholz} pointed out that the formal errors were very small and were due to the few number of epochs and so that the errors could be unrealistic.  The estimated errors by \citet{Scholz} were  24~mas/yr, which was the scatter of their data around the best fit to the model.

\citet{Burningham2013}  presented  YJHK spectroscopy for this object using the Gemini Near Infrared Spectrograph (GNIRS) and derived a type a spectral type of $T5 \pm 0.5$.

Comparing the results from \citet{Scholz} for the proper motions with our results and taking the errors into account, the results are compatible. 
With regard to  Fig.~\ref{W1-W2_SpT}, we can conclude that this object is slightly  overluminous in WISE by around 0.2~mag. In Fig.~\ref{J-H_MJ_models}, this overluminosity is not that clear, although the object is at the edge of the bulk of objects. Our result for the distance is  $30.3_{-3.8}^{+2.5}$~pc.

\subsection{2MASS J22541892+3123498}\label{J2254}

The object 2MASS~J22541892+3123498 was first published as a brown dwarf by \citet{Burgasser}. It was classified as a T5V BD by using spectral observations in NIR with R $\sim$ 1200 and by calculating the spectral indices, as explained in \citet{Burgasser}. Afterward, the spectral type was recalculated using the criteria given in \citet{Burgasser2003} as a standard T4 BD \citep{Burgasser2004, Burgasser2006}. The proper motions were also estimated by \citet{Jameson}  using WFCAM at UKIRT from February to August 2006. Their result for the proper motions was $\mu_{\alpha}\cos{\delta} = 68 \pm 15$ ~mas/yr and $\mu_{\delta} = 200\pm 11$~mas/yr.

Our results for the proper motions agree with the result given by \citet{Jameson} within the error bars. This object is not overluminous. The provided distance is $13.9_{-0.6}^{+0.5} $~pc, and it has been calculated for the first time.

\subsection{ULAS J232035.28+144829.8}\label{J2320}

The discovery of this object was first published by \citet{Burningham2},  and it was classified as a T5 spectral type using medium resolution spectroscopy. In \citet{Scholz}, the object is classified as T7.0$\pm$2.0  using colors from UKIDSS and SDSS. 
In \citet{Murray}, the spectral indices were recalculated, using data from NIRI on the Gemini-North telescope with a spectral resolution of R$\sim$460; the new classification is T6$\pm$1.

In \citet{Scholz}, the proper motions were given as $\mu_{\alpha}\cos{\delta} = 387 \pm 5$~mas/yr  and $\mu_{\delta} = 121 \pm 2$~mas/yr. These proper motions were calculated  with a baseline of seven years; nevertheless, there were only three epochs in total, and two of them very close to each other. The estimated error in \citet{Scholz} due to the scatter around  the best fit to the model was 11~mas/yr.  

In \citet{Murray}, the proper motions and the distance were also estimated performing a photometric follow-up using UKIDSS, by obtaining   proper motions of $\mu_{\alpha}\cos{\delta} = 399 \pm 26$~mas/yr, $\mu_{\delta} = 122 \pm 26$~mas/yr, and distance of $24\pm5$~pc.

The proper motions, which we obtained, agree to  the values provided by \citet{Scholz} and \citet{Murray} within the error bars, but our values are more precise. We also agree with \citet{Murray} on the distance to within the error bars. This object is not overluminous (see Fig. \ref{J-H_MJ_models} and \ref{W1-W2_SpT}). The calculated distance is $ 21.1_{-2.2}^{+1.6} $~pc.

\subsection{ULAS J232123.79+135454.9}\label{J2321}

The brown dwarf discovery of ULAS~J232123.79+135454.9 was first published by \citet{Scholz}. It was classified with a spectral type of T$7.5\pm1.5$ using UKIDSS and SDSS colors. This object was also classified in \citet{Burningham2} as a T7.5 using medium resolution spectroscopy.

\citet{Scholz} also  provided the proper motions  using SDSS and UKIDSS with a baseline of seven years and four epochs taken along this baseline. The proper motions were given as $\mu_{\alpha}\cos{\delta} = 56 \pm 15 $~mas/yr and $\mu_{\delta} = -577 \pm 10$~mas/yr  with an estimated error of 10~mas/yr due to the scatter of the data around the fit to the best model. \citet{Kirkpatrick2012} provided a  distance limit of approximately 20~pc. 

\begin{figure*}
\centering
\includegraphics[width=9cm]{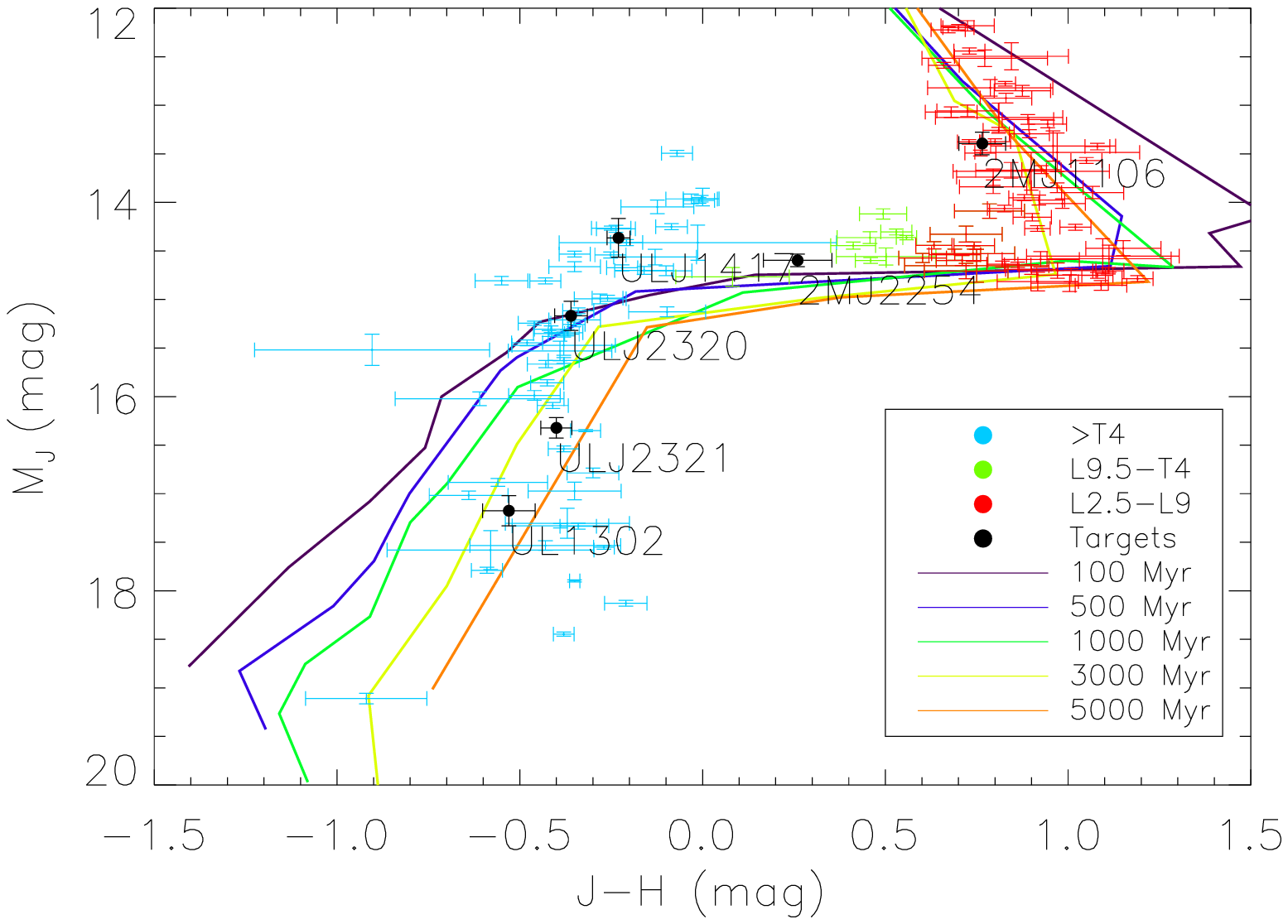}
\includegraphics[width=9cm]{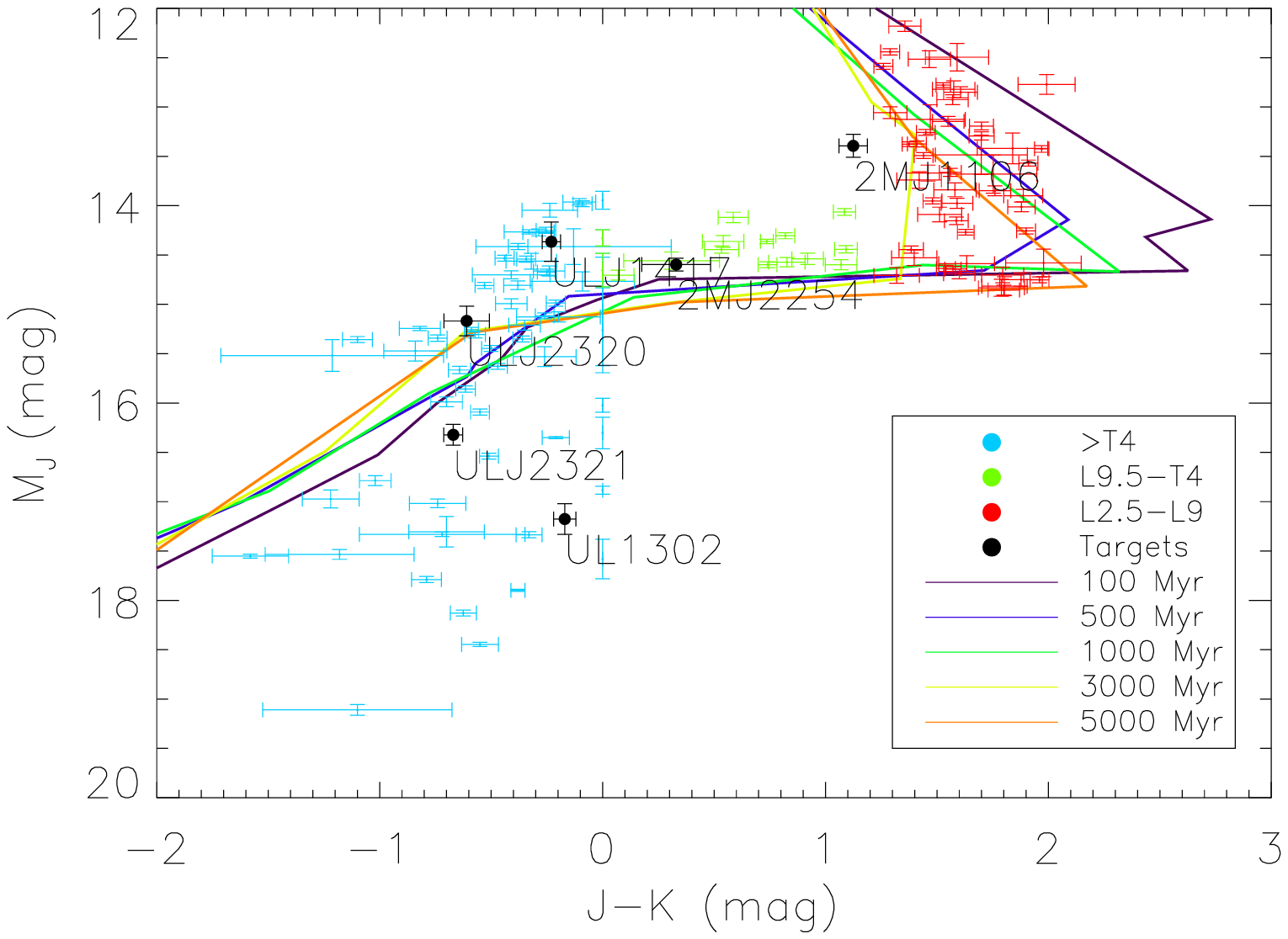}

  \caption{Color-magnitude diagram in the MKO system showing the brown dwarf sample from \citet{Dupuy} (except M-type brown dwarfs), our targets and the BT~Settl-models \citep{Allard}. The objects in the \citet{Dupuy} sample with spectral type between L2.5-L9 are shown in red; the objects with spectral types between L9.5 and T4 are shown in green; and the objects with spectral type >T4 are shown in light blue. Our targets, which are listed in Table \ref{targets}, are shown in black. We overplot the evolutionary models from \citet{Allard} in  MKO photometry for different ages from 100~Myr to 5000~Myr.}
  \label{J-H_MJ_models}
\end{figure*}

\begin{figure*}
\centering
	\includegraphics[width=10cm]{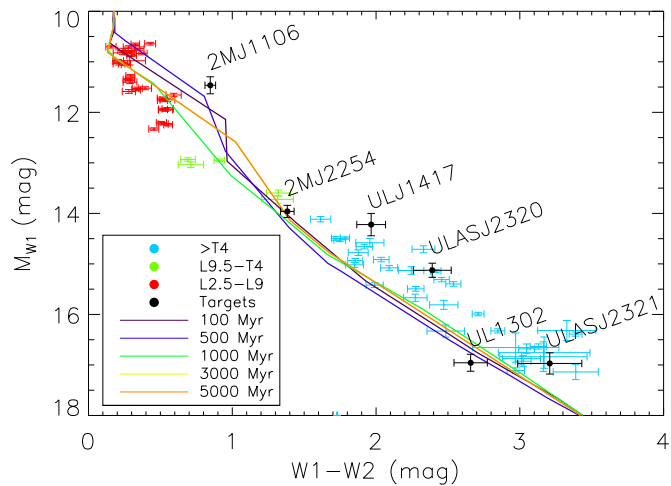}
	\caption{Color-magnitude diagram in WISE photometry with W1-W2 color vs $M_{W1}$ (WISE). We plot the objects from \citet{Dupuy} with WISE photometry. L2.5-L9 brown dwarfs are colored in red; objects with spectral types between L9.5 and T4 are green; and objects with spectral type >T4 are colored in light blue. Our targets are plotted in black. We overplot  the evolutionary models from \citet{Allard} in WISE photometry for different ages, from 100~Myr to 5000~Myr.}
\label{W1-W2_SpT}
\end{figure*}

The object ULAS~J232123.79+135454.9 is not overluminous  (see Fig. \ref{J-H_MJ_models} and \ref{W1-W2_SpT}). We agree with \citet{Scholz}   on the result for the proper motion within the error bars.  \citet{Kirkpatrick2012} adopted a distance of 14.1~pc. Our result for the distance is $11.8_{-0.6}^{+0.5}$~pc.

\begin{table*}
\caption{Summary of the results.}            
\label{parallaxes}      
\centering          
\begin{tabular}{l r r r r r r l}     
\hline\hline       
Object & $\mu_{\alpha}~$(mas/yr) &$\mu_{\delta}$~(mas/yr) &$\pi_{rel}$~(mas) & $\pi_{abs}$~(mas)& d~(pc) & $d_{phot}$~(pc) & $\chi^{2}$/$N_{dof}$ \\
 \hline

2M~ J1106+2754   & $-311\pm4$ & $-438\pm5$ &$ 46\pm3$  & $48\pm3 $ & $20.6_{-1.2}^{+1.0}$   &$ 12.5\pm1.4$   & $21.8/17$ \\
ULAS~ J1302+1308 & $-445\pm6$ &$ 5\pm7 $   & $67\pm5$  & $65\pm5$  & $15.4_{-1.4}^{+1.1}$   & $16.1\pm2.3$   & $25.4/19$ \\
ULAS~ J1417+1330 & $-121\pm4$ &$ 50\pm3 $  & $32\pm3$  & $33\pm3 $ & $30.3_{-3.8}^{+2.5}$   & $23.8\pm5.1$   & $23.0/17$ \\
2M~ J2254+3123   & $67\pm3$   &$ 187\pm7 $ & $71\pm2$  & $72\pm3$  & $13.9_{-0.6}^{+0.5} $ & $14.7\pm1.9$   & $23.4/17$\\
ULAS ~J2320+1448 & $410\pm4$  &$ 121\pm3 $ & $47\pm3$  & $47\pm4$  & $ 21.1_{-2.2}^{+1.6} $ & $20.7\pm3.6$   & $22.6/19$\\
ULAS~ J2321+1354 & $76\pm4$   &$ -576\pm6 $& $83\pm3$  & $84\pm4$  & $11.8_{-0.6}^{+0.5}$   & $10.8\pm0.7$   & $22.2/19$ \\

\hline                  
\end{tabular}
\end{table*}


\section{Conclusions}\label{conclusions}

We have measured the trigonometric parallaxes of six T brown dwarfs for the first time with spectral types between T2.5 and T7.5. 

We compare our results to the spectrophotometric distance given by \citet{Goldman} and to the results of the spectrophotometric distance and proper motions provided in other studies such as \citet{Scholz}, \citet{Looper} and \citet{Kirkpatrick2012} among others. Our results generally agree well with other studies, but they are more precise.

We also compare the locations of our targets in (J-H, $M_{J}$) and (J-K, $M_{J}$) CMDs to those of the objects by  \citet{Dupuy} and to the evolutionary tracks by  \citet{Allard}.

Four of our six targets are not overluminous. Nevertheless, the object  ULAS~J141756.22+133045.8 is slightly overluminous in the WISE CMD. It has an absolute magnitude in W1, which is around 0.2~mag brighter than the objects with the same spectral types. Its overluminosity is not seen in the CMDs in MKO photometry. The object 2MASS~J11061197+2754225 is more than 1~mag overluminous in all the bands, pointing to binarity or even higher multiplicity. To confirm these results,  high resolution imaging and high resolution spectroscopy is needed.


\begin{acknowledgements}

We acknowledge J. A. Caballero, M. R. Zapatero Osorio, V. J. S. B\'ejar,  W. Brandner, A. Bayo, A. Burgasser and B. Burningham for their help, their corrections and contructive criticism. We thank  in particular to our referee T. Dupuy for his useful advice, which substantially helped to improve our paper.
This work was supported by Sonderforschungsbereich SFB 881 \textit{The Milky Way System} (subproject B6) of the German Research Foundation (DFG).
      
\end{acknowledgements}

\bibliographystyle{aa}
\bibliography{parallax.bib}

\begin{appendix}\label{appendix}
\section{Stellar paths of the targets}

We include the stellar paths, which represent the change in right ascension (RA) and declination (DEC) in time for our targets in this Appendix. 

\begin{figure*}[h!]
\sidecaption
	\includegraphics[width=.5\textwidth]{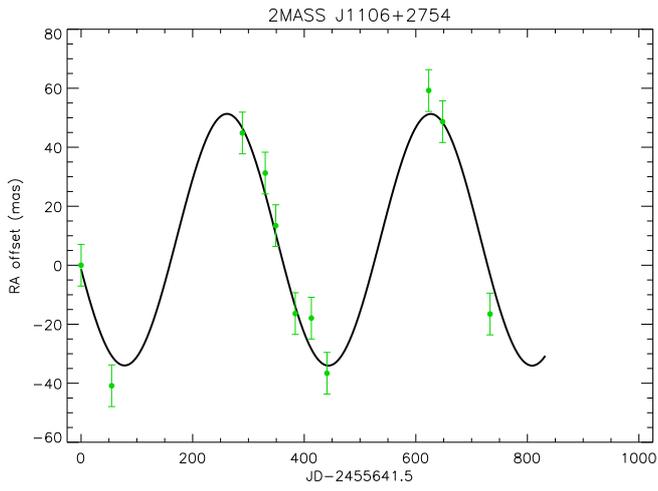}
	\includegraphics[width=.5\textwidth]{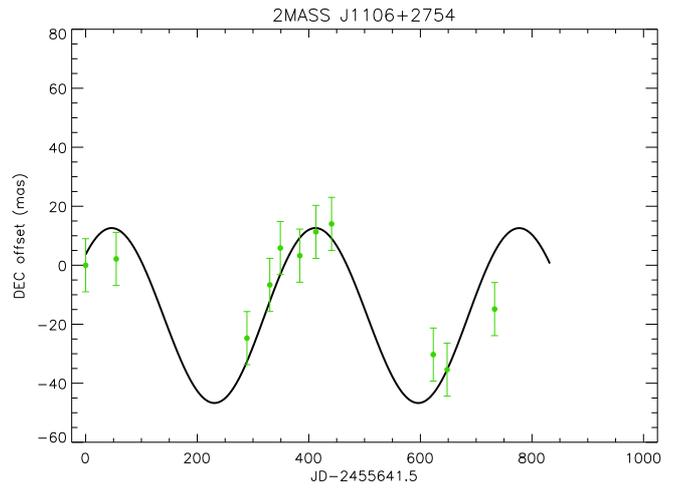}
	\caption{Stellar paths for the object 2MASS~J11061197+2754225.}
\label{dispersionRA2j_dispersionDEC2j}
\end{figure*}


\begin{figure*}[h!]
\sidecaption
	\includegraphics[width=.5\textwidth]{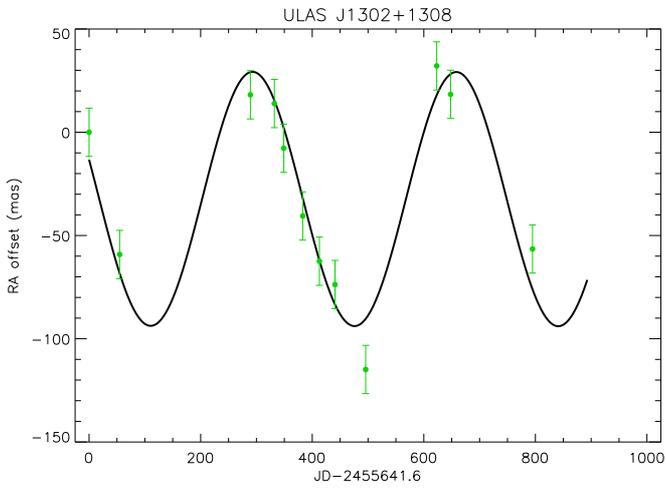}
	\includegraphics[width=.5\textwidth]{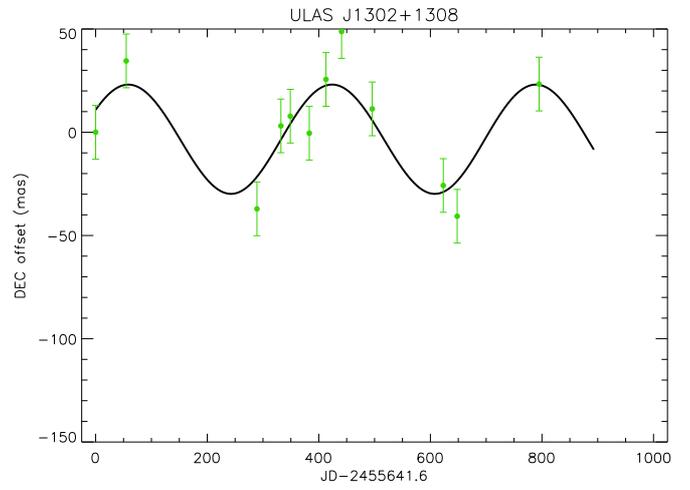}
	\caption{Stellar paths for the object ULAS~J130217.2+130851.2.}
\label{dispersionRA3j_dispersionDEC3j}
\end{figure*}


\begin{figure*}[h!]
\sidecaption
	\includegraphics[width=.5\textwidth]{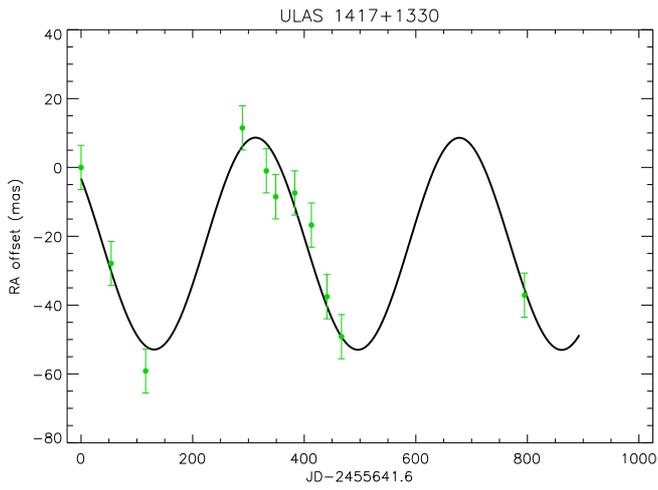}
	\includegraphics[width=.5\textwidth]{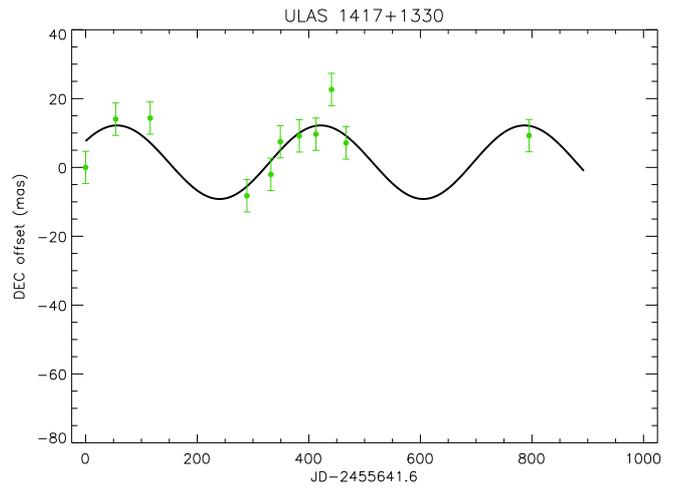}
	\caption{Stellar paths for the object ULAS~J141756.22+133045.8.}
\label{dispersionRA4j_dispersionDEC4j}
\end{figure*}


\begin{figure*}[h!]
\sidecaption
	\includegraphics[width=.5\textwidth]{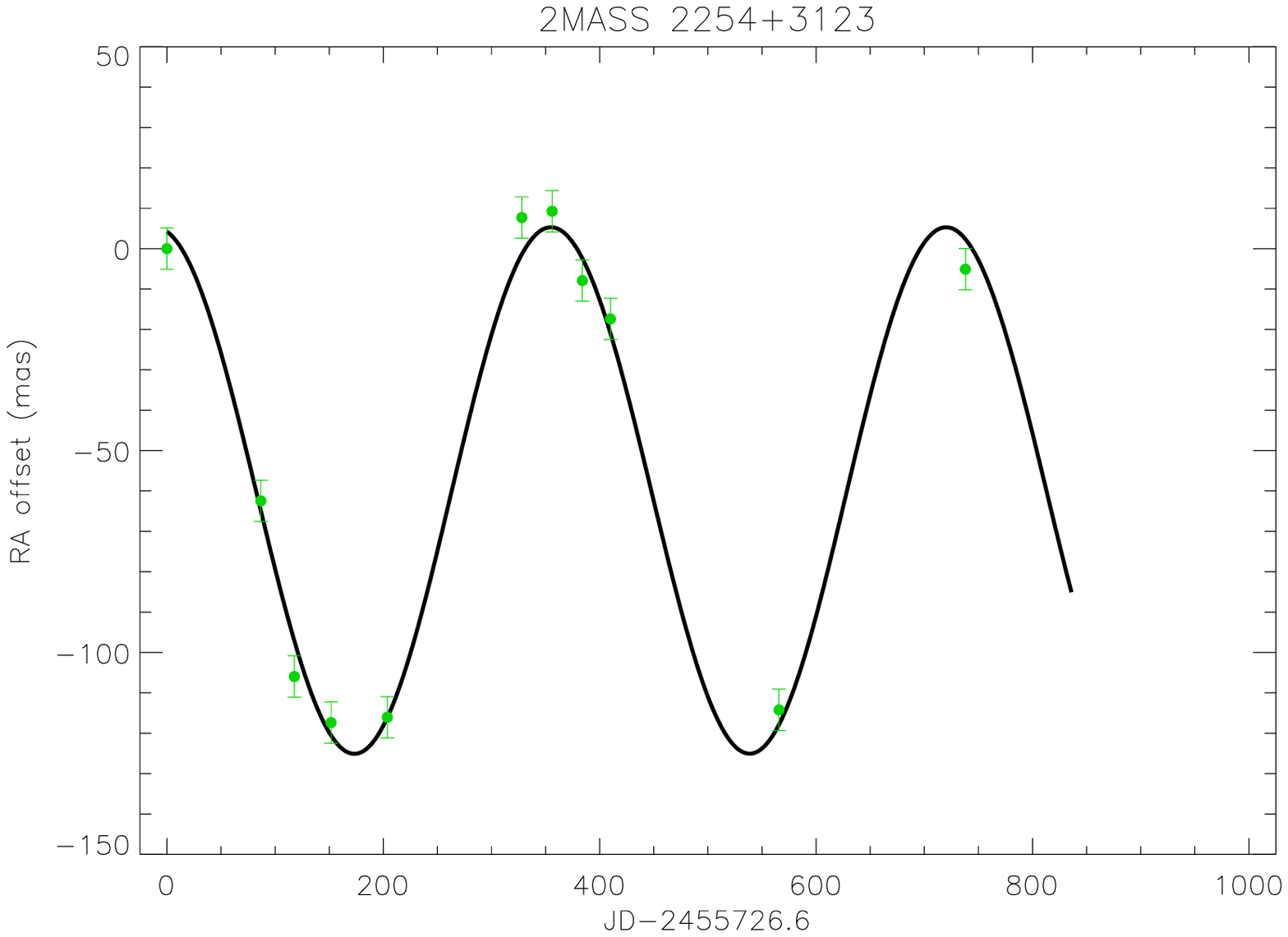}
	\includegraphics[width=.5\textwidth]{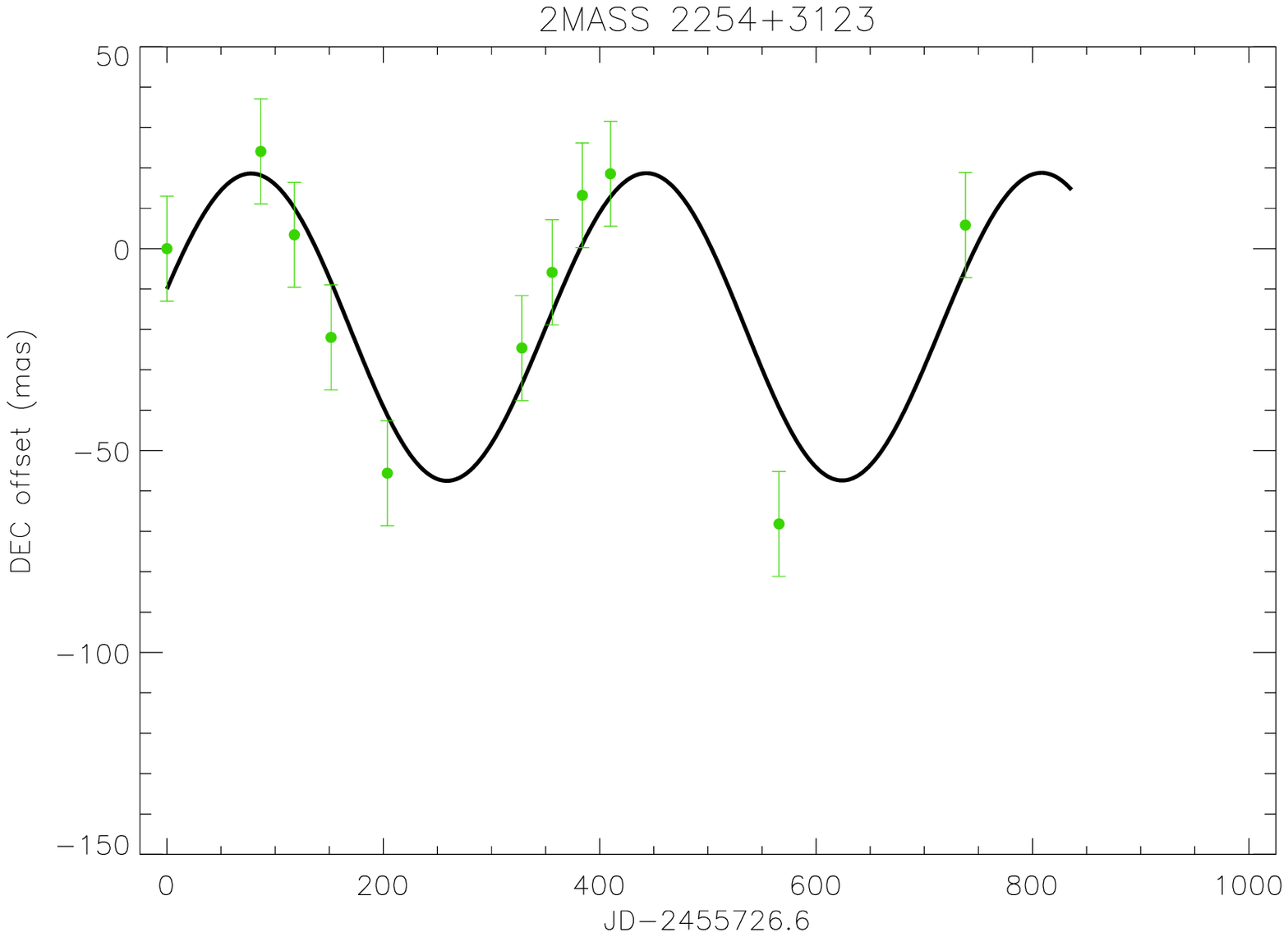}
	\caption{Stellar paths for the object 2MASS~J22541892+3123498.}
\label{dispersionRA6j_dispersionDEC6j}
\end{figure*}


\begin{figure*}[h!]
\sidecaption
	\includegraphics[width=.5\textwidth]{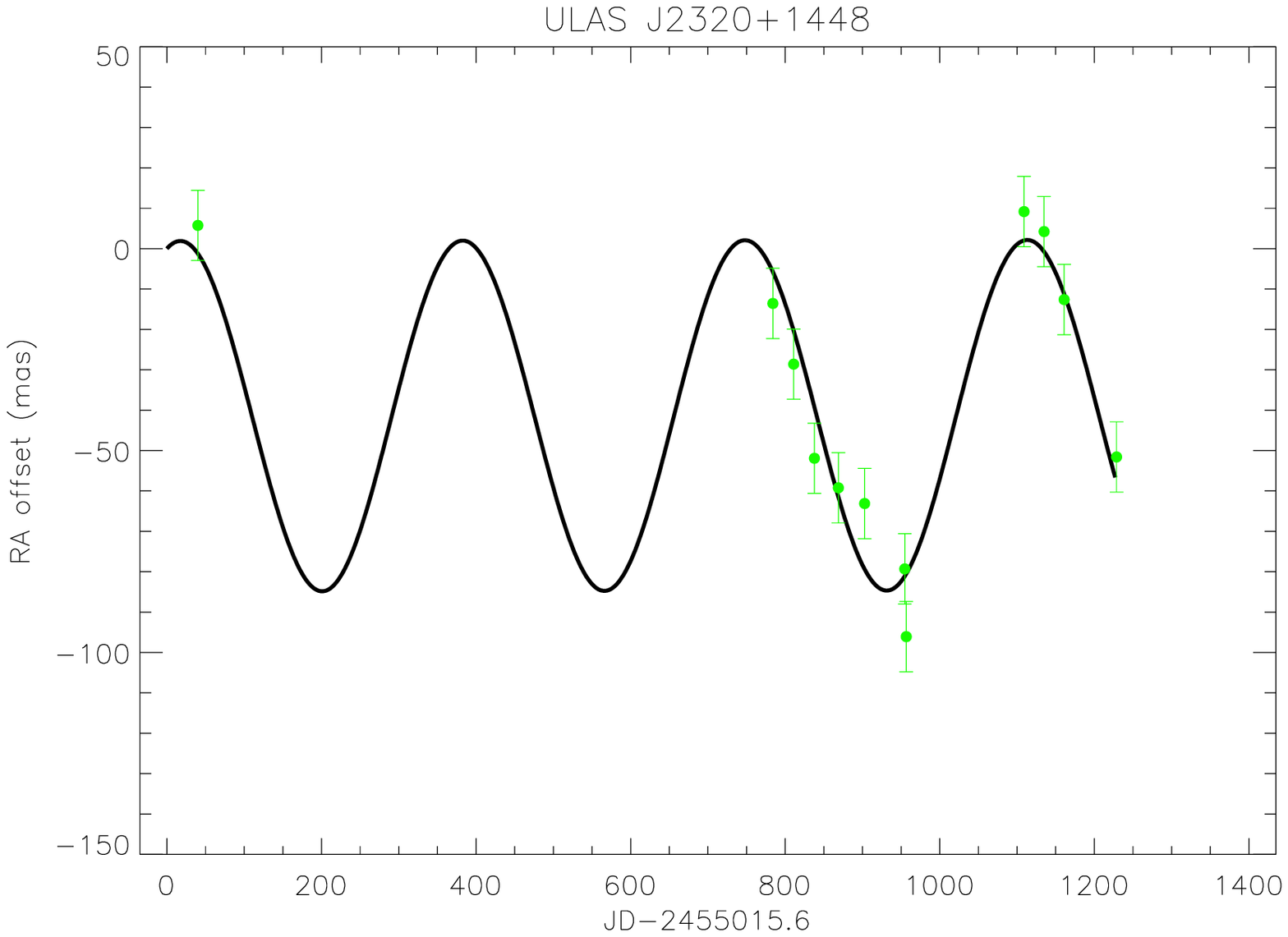}
	\includegraphics[width=.5\textwidth]{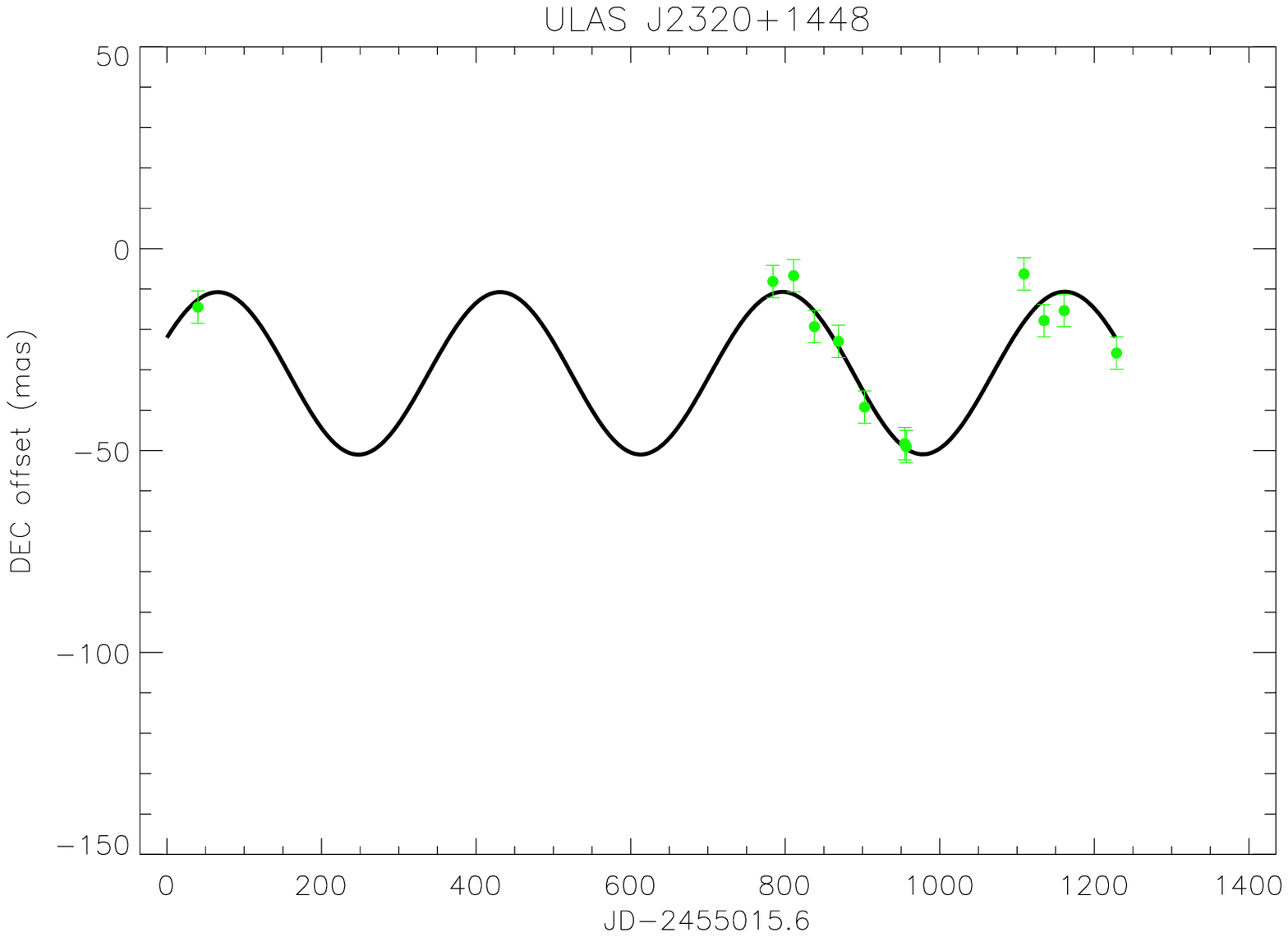}
	\caption{Stellar paths for the object ULAS~J232035.28+144829.8. The first epoch is an archival Omega-2000 observation.}
\label{dispersionRA7j_old_dispersionDEC7j_old}
\end{figure*}


\begin{figure*}[h!]
\sidecaption
	\includegraphics[width=.5\textwidth]{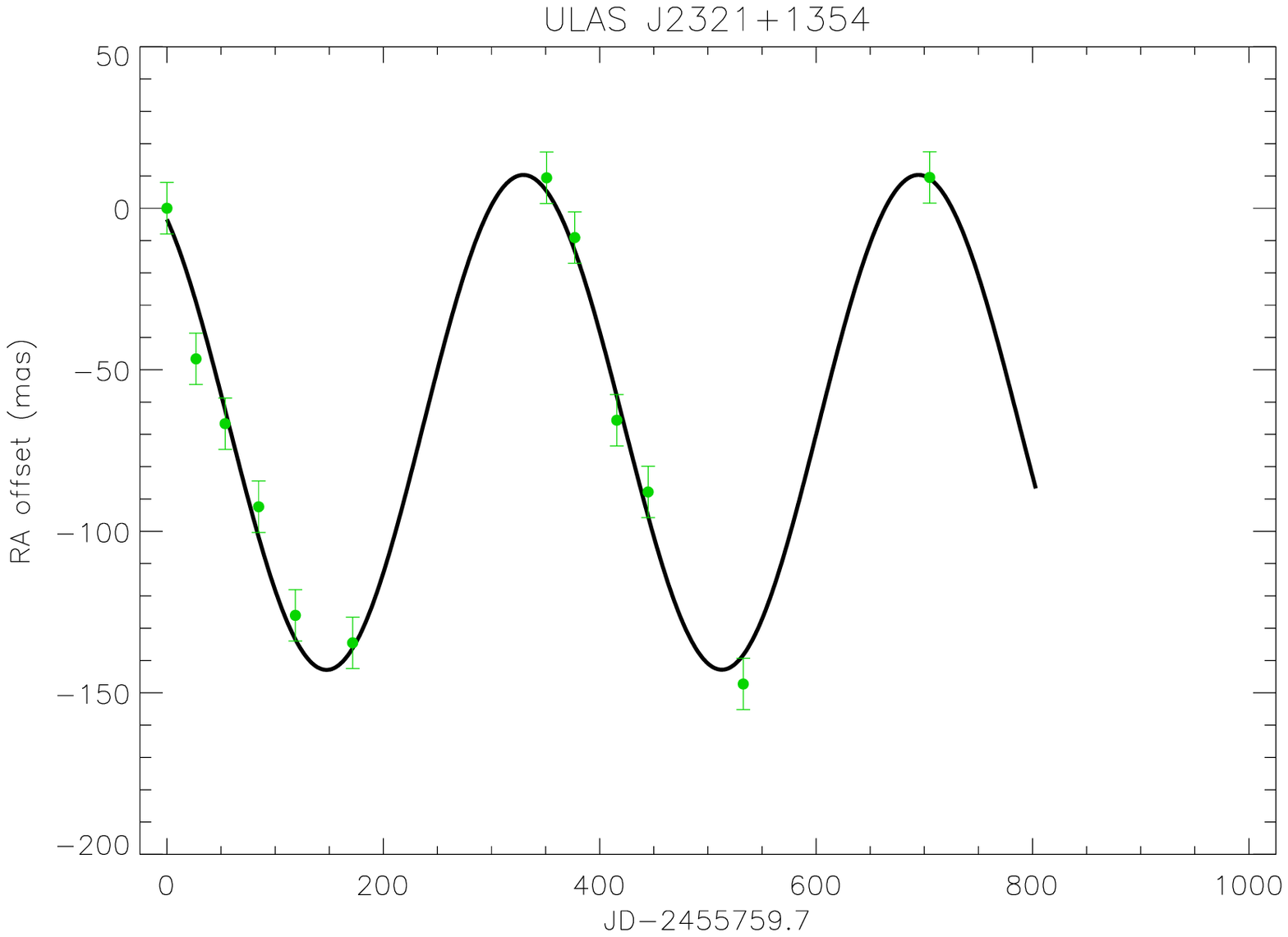}
	\includegraphics[width=.5\textwidth]{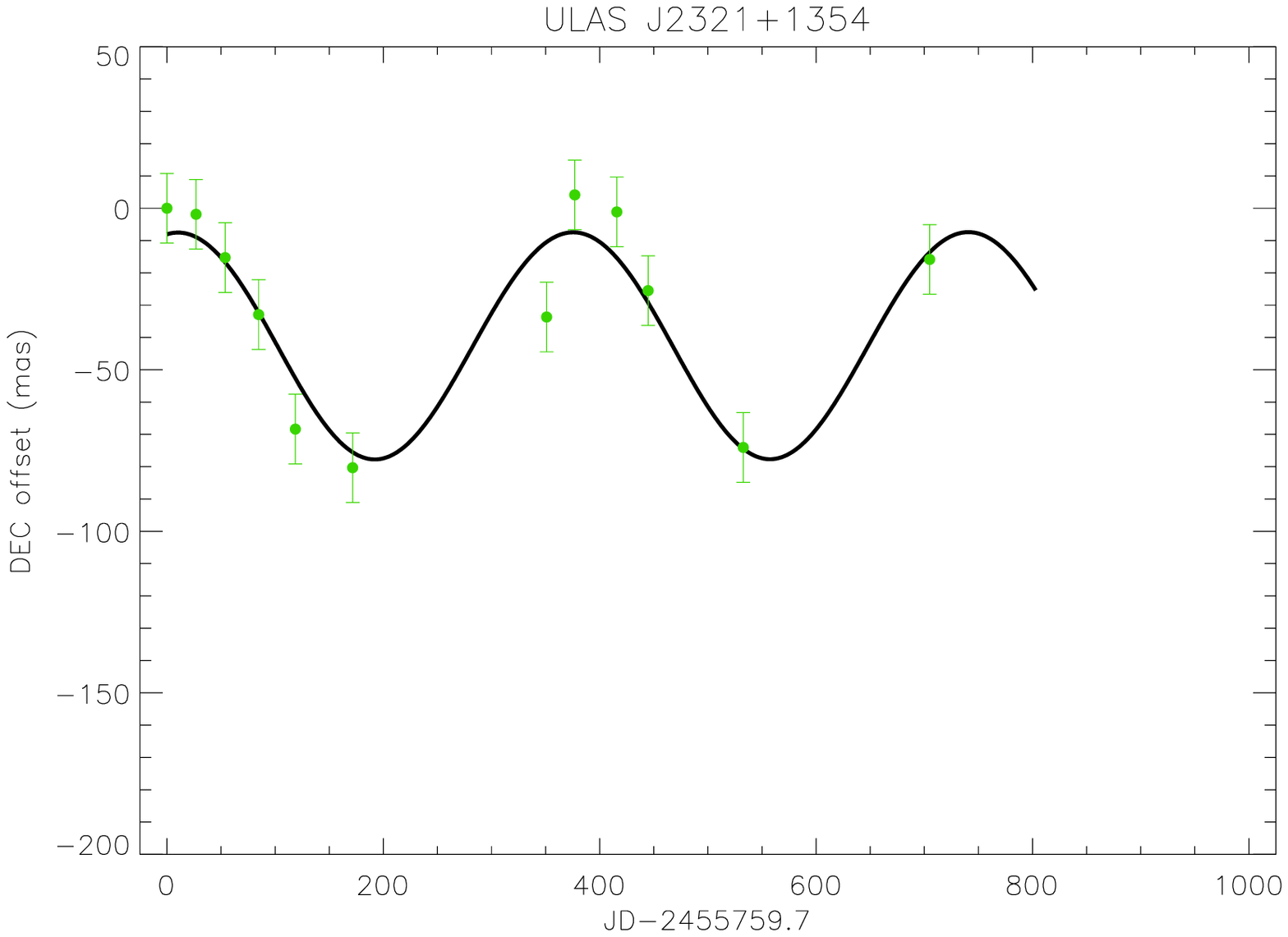}
	\caption{Stellar paths for the object ULAS~J232123.7+1354454.}
\label{dispersionRA8j_dispersionDEC8j}
\end{figure*}

\end{appendix}

\end{document}